\documentclass[twocolumn, 
prb,preprintnumbers,amsmath,amssymb,floatfix]{revtex4}

\usepackage{graphicx}
\usepackage{subfigure}
\usepackage{dcolumn}

\begin{document}

\title{Stability of the Gd Magnetic Moment to the 500 GPa Regime}
\author{Z. P. Yin and W. E. Pickett}
\affiliation{Department of Physics, University of California, 
    Davis, California, 95616}

\date{\today}
%%%%%%%
\begin{abstract}
The evolution of the magnetic moment and various features of the
electronic structure of fcc Gd are followed to reduced volume
$V/V_o = 0.125$ using the LDA+U correlated band method.  The
stability of the moment is substantial; crude estimates of
this signature of a possible ``Mott transition'' in the $4f$
system suggest a critical pressure $P_c \sim$ 500 GPa.  The
$4f$ occupation is found to {\it increase} under pressure due
to broadening and lowering of the minority states.  This trend
is consistent with the interpretation of x-ray spectra of
Maddox {\it et al.} across the volume collapse transition at
59 GPa, and tends to support their suggestion that
the delocalization of the $4f$ states in Gd differs from the original
abrupt picture, being instead a process that occurs
over an extended range of pressure.
\end{abstract}
\maketitle

\section{Introduction}
The behavior of the $4f$ rare earth metals and their compounds under
pressure has been discussed for decades, with the volume collapse
transition under pressure attracting a great deal of attention.
It has been known for some time that there
are volume collapse transitions in Ce (15\% at 0.7 GPa), Pr (10\%
at 20 GPa), Gd (5\% at 59 GPa) and Dy (6\% at 73 GPa), while no
significant volume collapses have been detected in
Nd, Pm, and Sm.  The equation of state of these metals, and references
to the original work, has been
collected by McMahan {\it et al.}\cite{andy1}
High temperature experiments\cite{err} have seen signatures
that are likely related to the localized$\rightarrow$itinerant
transition, at 50 GPa in Nd and 70 GPa in Sm.

The question can be stated more generally as: what form does the 
localized $\to$ itinerant transition of the $4f$ states take, and
what is the correct description?  This transition is intimately related
to the question of behavior of magnetic moments,\cite{andy1} although the 
questions are not the same.  There have been two main viewpoints on the
volume collapse transition. One is
the ``Mott transition of the $4f$ system'' elaborated by 
Johansson,\cite{borje} in which the crucial ingredient is the change
from localized (nonbonding) to more extended states (participating
in bonding), with an accompanying drop in magnetic tendency.  The
other is the ``Kondo volume collapse'' view introduced by Allen and
Martin\cite{allen} and Lavagna {\it et al.}\cite{lavagna}, in which the main
feature is the loss of Kondo screening of the local moment, with
a decrease in localization of the $4f$ state not being an essential
feature.

At ambient conditions the
$4f$ electrons form a strongly localized $f^n$ configuration that is
well characterized by Hund's rules.  Under reduction of volume,
several things might be anticipated to happen.  At some point the $4f$
system begins to respond to the non-spherical environment. 
Initially, perhaps, it is just a matter of crystal field splitting
becoming larger.  Then the $4f$ orbitals actually begin to become
involved in the electronic structure, by overlapping orbitals of
neighboring atoms.  The consequences of this are possible participation
in bonding, and that the orbital moment becomes less well defined
(the beginning of quenching {\it i.e.} the loss of Hund's second
rule, which has already occurred in magnetic
$3d$ systems).  Additionally, the $4f$ levels can shift and increase
their interaction with the itinerant conduction ($c$) bands (Kondo-like
coupling), which can change the 
many-body behavior of the coupled $4f-c$ system.  At some point 
the kinetic energy
increase, characterized by the $4f$ bandwidth $W_f$, compared to the 
on-site interaction energy $U_f$ reaches a point
where the spin moment begins to decrease.  Finally, at small enough
volume (large enough $W_f$) the $4f$ states simply form nonmagnetic
conduction bands.

Just how these various changes occur, and in what order and at what
volume reduction, is being addressed in more detail by recent high
pressure experiments.  
Here we revisit the case of
Gd, whose volume collapse was reported by Hua {\it et al.}\cite{hua} 
and equation of state by Akella {\it et al.}\cite{akella}
The deviation from the series of close-packed structures below
P$_c$=59 GPa and the lower symmetry bcm (body-centered monoclinic)
high pressure structure signaled the expected onset of 
$f$-electron participation
in the bonding, and Hua {\it et al.} seemed to expect that the 
moment reduction and delocalization of the $4f$ states would
accompany this collapse.

New information has been reported by Maddox {\it et al.},\cite{maddox} 
who have monitored
the resonant inelastic x-ray scattering and x-ray emission
spectra  of Gd through P$_c$ and up to 113 GPa.  They find that
there is no detectable reduction in the magnetic moment at the volume 
collapse transition, so the volume-collapse is only a part of a
more complex and more extended
delocalization process of the $4f$ states.  Maddox {\it et al}
emphasize the Kondo-volume-collapse\cite{kondo1,kondo2} aspects of the 
transition at P$_c$.

The treatment of the $4f$ shell, and particularly the volume collapse
and other phenomena that may arise (see above), comprises a
correlated-electron problem for theorists.  Indeed there has been
progress in treating this volume-collapse, moment-reduction
problem in the past few 
years.  The issue of the (in)stability of the local moment seems to involve 
primarily the local physics, involving the treatment of the hybridization 
with the conduction bands and interatomic $f-f$ interaction, 
with Kondo screening of the moment
being the subsequent step.
Dynamical mean field studies of the full multiband 
system have been carried out for Ce\cite{andy1,held,andy2} and for Pr
and Nd.\cite{andy2}  These calculations were based on a well-defined
free-energy functional and included the conduction bands as well as
the correlated $4f$ bands.  One simplification was that only an
orbital-independent Coulomb interaction $U$ was treated, leaving the
full orbital-dependent interaction (fully anisotropic Hund's rules) for
the future.  Density functional based correlated band theories have 
also been applied (at zero temperature).  Self-interaction corrected 
local density approximation (LDA) was applied to Ce, obtaining a
volume collapse comparable to the observed one.\cite{SIC}  Four correlated
band theories have been applied\cite{MnO} to the antiferromagnetic insulator
MnO.  Although their predictions for critical pressures and amount
of volume collapse differed, all obtained as an S=5/2 to S=1/2 moment
collapse rather than a collapse to a nonmagnetic phase.

Clearly there
remain fundamental questions about how the magnetic moment
in a multielectron atom 
disintegrates as the volume is reduced: catastrophically, to an
unpolarized state, or sequentially, through individual spin flips
or orbital-selective delocalization.  If the latter, the 
total (spin + orbital) moment
could actually {\it increase} initially in Gd.  If the occupation
change is toward $f^8$, the decrease in spin moment (from S=7/2 to S=3) could
be more than compensated by an orbital moment (L=3).  If the
change is toward $f^6$, the onset of an L=3 orbital configuration could
oppose the S=3 spin (Hund's third rule), leaving a non-magnetic J=0
ion (as in Eu$^{3+}$) even though the $4f$ orbitals are still localized.
Still another scenario would be that the increasing crystal field
quenches the orbital moment (as in transition metals) and the remaining problem
involves only the spins.

Our objective here is to look more closely at the stability of the Gd
atomic moment, in the general context of the localized$\to$itinerant
transition of the $4f$ system under pressure.  Consideration of the changes
in electronic structure under pressure go back at least to the broad study
of Johansson and Rosengren\cite{johansson} but most have not considered
the magnetism in detail.
We apply the LDA+U (local density approximation plus
Hubbard U) method to study the
evolution of the electronic structure and magnetism
as the volume is reduced.  Although this correlated band method
neglects fluctuations and the dynamical interaction with the
conduction electrons, it does treat the full multiorbital system
in the midst of itinerant conduction bands.
The resulting moment vs. volume surely provides only an
upper limit to the pressure where the moment decreases rapidly.  However,
we can invoke studies of the insulator-to-metal transition in multiband
Hubbard models to provide a more realistic guideline on when the 
localized$\to$itinerant (or at least the reduction in moment 
transition within the $4f$ system may be expected
to occur.  The results suggest stability of the moment to roughly
the 500 GPa region.

\section{Electronic Structure Methods}
In the paper we apply the full potential local orbital code\cite{FPLO}
(FPLO5.00-18) to Gd from ambient pressure to very high pressure (a few TPa).
We use the fcc structure with space group 
Fm3m (\#225) and ambient pressure atomic volume (corresponding to the fcc lattice 
constant a$_0$=5.097\AA). 
The basis set
is (core)::$(4d4f5s5p)/6s6p5d+$. 
We use 48$^3$ k point mesh and Perdew and Wang's PW92 functional\cite{PW92}
for exchange and correlation. 
We have tried both 5.0 and 6.0 for the confining potential exponent, with very
similar results, so
only the results using exponent=5 will be presented here.
We perform both LDA and LDA+U
calculations (see below).  Due to the extreme reduction in volume that
we explore, any band structure method might encounter difficulties.  For
this reason we have compared the FPLO results on
many occasions with parallel calculations with the 
full potential linearized augmented 
plane wave method WIEN2k.\cite{WIEN2k}  The results compared very well
down to $V/V_o$=0.5, beyond which the WIEN2k code became more difficult
to apply.
We use the notation $v\equiv V/V_o$ for the
specific volume throughout the paper.

We assume ferromagnetic ordering in all calculations.  The Curie temperature
has been measured only to 6 GPa,\cite{robinson,jackson1,jackson2} where it
has dropped from 293 K (P=0) to around 210 K.  Linear extrapolation suggests
the Curie temperature will drop to zero somewhat below 20 GPa.  However, as the
$4f$ bands broaden at reduced volume the physics will change substantially,
from RKKY coupling at ambient pressure finally to band magnetism at very high
pressure.  Antiferromagnetic (AFM) ordering does not affect the $4f$
bandwidth\cite{kurz} until $f-f$ overlap becomes appreciable.  AFM ordering
might affect some of the quantities
that we look at in this study at very high pressure, but such
effects lie beyond the scope of
our present intentions.

\subsection{LDA+U Method}
For the strength of the $4f$ interaction we have used the
volume dependent $U(V)$ calculated by McMahan {\it et al.},\cite{andy1}
which is shown below.  Due to the localized $4f$ orbital and the large
atomic moment, we use the ``fully localized limit'' version of
LDA+U as implemented in the linearized augmented planewave 
method,\cite{sasha} and as usual the ratio of Slater integrals is fixed at
$F_4/F_2$=0.688, $F_6/F_2$=0.495.  Since we are particularly interested
in the stability of the atomic moment, the exchange integral
$J$ that enters the LDA+U method deserves attention.
In atomic physics, and in the LDA+U method, the exchange 
integral plays two roles.  It describes the 
spin dependence of the Coulomb interaction, that is, the usual
Hund's rule coupling.  In addition, it carries the orbital
off-diagonality; with $J$=0 all $4f$ orbitals repel
equally by $U$, whereas in general the anisotropy of the 
orbitals leads to a variation\cite{EuN} that is described by $J$.

For a half filled shell for which the orbital occupations
$n_{m\uparrow}=1$ and
$n_{m\downarrow}$=0 for all suborbitals $m$, 
the exchange effect primarily counteracts
the effect of $U$, since the anisotropy of the repulsion averages
out.  As a result, using $U_{eff} \equiv U - J$ with $J_{eff}$=0
is almost equivalent, for a perfectly half-filled shell, 
to using $U$ and $J$ separately as normally is done.
Since it could be argued that Hund's first rule is treated adequately
by the LDA exchange-correlation functional, for our calculations 
we have set $J$=0.  This becomes approximate for the off-diagonality
effects
when the minority $4f$ states begin to become occupied at 
high pressure.  However, we have checked the effect
at $a/a_o$=0.8 ($v$
= 0.5, $P$=60 GPa).
Comparing $U=6.9$ eV, $J$=1 eV with $U=5.9$ eV, $J$=0, we find the
energy is exactly the same (to sub-mRy level) and the moment is
unchanged.  This result is in line with the $U_{eff}, J_{eff}$ 
argument mentioned above.
Changing $J$ from 1 eV to 0 with $U=5.9$ eV also leaves the energy
unchanged, illustrating the clear unimportance of $J$.  The $J$=1 eV 
calculation does result in a 0.03 $\mu_B$ larger moment.
At much smaller volumes, where the minority bands overlap the Fermi
level, the changes become noticeable and would affect the equation
of state, but only in a very minor way. 
In general, neglect of $J$ will tend to
{\it underestimate} the stability of the magnetic moment, which we
show below already to be extremely stable.

\subsection{Structure}
The observed structures of Gd follow the sequence 
hcp$\rightarrow$Sm-type$\rightarrow$dhcp$\rightarrow$dfcc
$\rightarrow$bcm 
(dfcc$\equiv$distorted fcc, which is trigonal; 
bcm$\equiv$body-centered monoclinic).  All except the bcm phase
are close-packed arrangements, differing only in the stacking of
hexagonal layers.  The bcm phase is a lower symmetry phase that
suggests $f$-electron bonding has begun to contribute.

For our purpose of studying trends relating only to the atomic 
volume, it is best to stay within a single crystal structure.
We expect the results to reflect mostly local physics,  
depending strongly on the volume but only weakly 
on the long-range periodicity.
Therefore we have kept the simple fcc structure for the results
we present.

\section{LDA+U Results}
The overall result of our study is that evolution of the volume
and the Gd moment are predicted by LDA+U to be continuous 
under reduction of volume, with
no evidence of a volume-collapse transition (or any other
electronic phase change) in the region where one
is observed (59 GPa), or even to much higher pressure.  
This result provides some support for the suggestion
that the volume collapse is Kondo-driven, or involves in an essential
way fluctuations, neither of which are accounted for in our approach.

\begin{figure}[tbp]
%\rotatebox{-90}
{\resizebox{7.2cm}{7.2cm}{\includegraphics{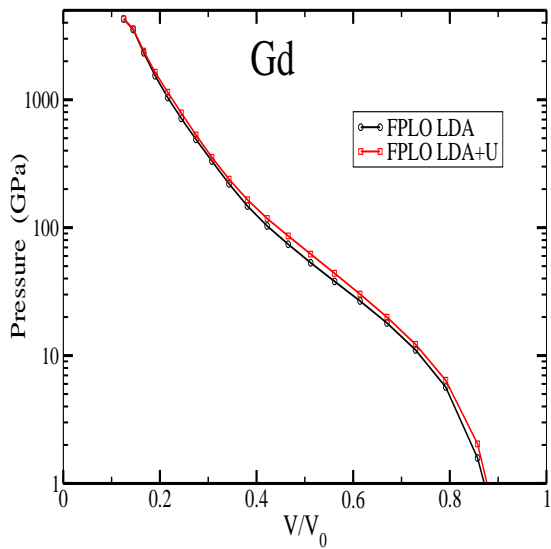}}}
\caption{
Log plot of the calculated pressure versus volume. The relatively
small difference between the LDA+U and LDA results is evident.  The 
relation is roughly exponential below $V/V_o < 0.8.$
Current static diamond anvil cells will only take Gd to
the $V/V_o \sim 0.35$ region.
 }
\label{logPV}
\end{figure}

While we will usually quote volumes or the relative volume $v$, 
it is useful to be able to convert 
this at least roughly to pressure.  We provide in Fig. \ref{logPV}
the calculated equation of state, plotted as log $P$ vs. $V/V_o$.
It can be seen that the pressure is very roughly exponential in -$V/V_o$
from $v$=0.8 down to $v$=0.15 (2 GPa to 4 TPa).  The change in 
slope around $v$=0.4 (in the vicinity of 100-200 GPa) is discussed below.

\subsection{Magnetic Moment vs. Volume}
\begin{figure}[tbp]
%\rotatebox{90}
{\resizebox{7.2cm}{7.2cm}{\includegraphics{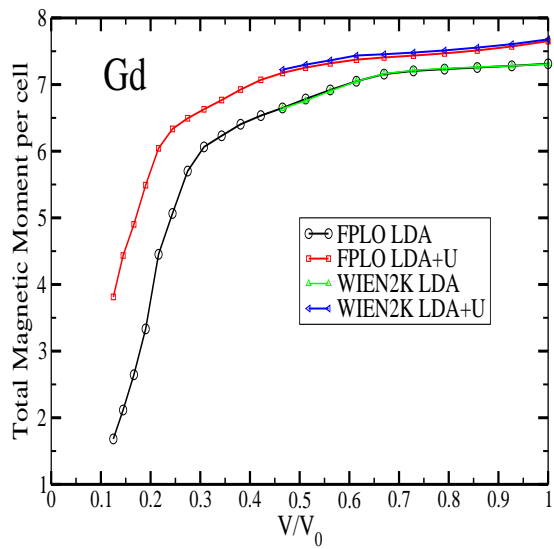}}}
\caption{(Color online)
Behavior of the calculated moment/cell ($4f$ spin moment plus
conduction electron polarization) of Gd versus reduction in  
volume, from both LDA and LDA+U methods. For the more realistic
LDA+U method, there is very little decrease in moment down to
$V/V_o$=0.45 ($\sim$110 GPa), with a rapid decline beginning
only around $V/V_o \approx$ 0.2 (1.5 TPa).
   }
\label{moment}
\end{figure}

The behavior of the total spin moment ($4f$ plus conduction)
in LDA+U is compared in Fig. \ref{moment}
with that of LDA.  The general trend is similar, but the decrease in
moment is extended to smaller volume by the correlations in LDA+U.
Specifically, the moment is reduced not by decrease of majority
spin population (which would be $f^7\rightarrow f^6$) but rather by
increase in the minority spin population ($f^7 \rightarrow f^8$;
see discussion below).  Thus LDA+U enhances the stability of the moment by
raising the unoccupied minority $4f$ states in energy, thus
reducing and delaying compensation of the filled majority states.
It has been noted elsewhere\cite{shick} that raising the minority states
is the main beneficial effect
of the LDA+U method for Gd at ambient pressure.

The decrease in moment is minor down to $v$ =
0.45 ($\sim$90-100 GPa)
beyond which the decrease from 7$\mu_B$ to 6$\mu_B$ occurs by $v$
= 0.2 ($P \sim$ 1 TGa).  Only beyond this incredibly high pressure does
the moment decrease more rapidly, as the $4f$ states become band-like.  
Even in LDA
this collapse does not occur until below $v$=0.3 ($P\sim$ 300-400 GPa).  
With the
neglect of fluctuations, the simplistic interpretation of the LDA+U
results is that the Gd ``bare'' spin moment
is relatively stable to $\sim$1 TPa.

It might be thought that, for the region of spin moment of 6 $\mu_B$
and below, where the minority occupation is one or more, there might
be an orbital moment of the minority system.  However, at these
volumes (see below) the minority $4f$ bandwidth is 5 eV or more, 
which we think makes an orbital moment unlikely.  Therefore
we have not pursued this possibility.

\subsection{4f Bandwidth}
\begin{figure}[tbp]
%\rotatebox{-90}
{\resizebox{7.2cm}{12.2cm}{\includegraphics{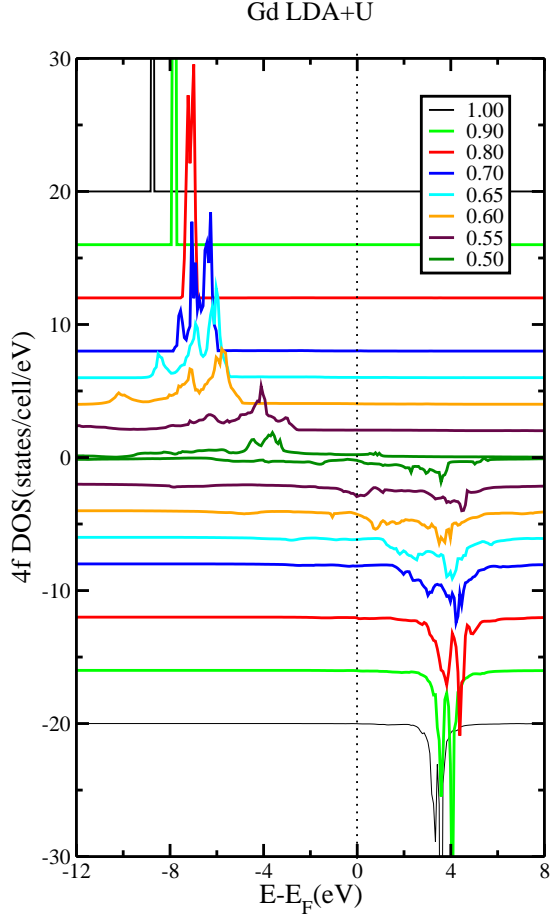}}}
\caption{(Color online)
View of the $4f$ projected density of states under compression,
with majority spin plotted upward and minority plotted downward.
The curves are displaced for clarity, by an amount proportional
to the reduction in lattice constant. The legend provides the
ratio $a/a_o$, which is decreasing from above, and from below,
toward the middle of the figure.
 }
\label{4fPDOS}
\end{figure}

The behavior of the $4f$ states, which become bands, is better
illustrated in Fig. \ref{4fPDOS}, where the evolution of the
$4f$ ``bands'' (the $4f$ projected density of states [PDOS])
is provided graphically.  At $a/a_o$=0.80
($v$=0.51, $P \approx$ 60 GPa, where the volume collapse is observed) 
the majority PDOS is somewhat less than 2 eV wide and still 
atomic-like, since it does not quite overlap the bottom of 
the conduction band.
Above this pressure range the $4f$ states begin to overlap the conduction
bands, primarily due to the broadening of the conduction bandwidth.
By $a/a_o$=0.70 ($v$=0.34, $P \approx$ 200 GPa) the width is at
least 3 eV and the shape shows the effect of hybridization and
formation of bands.  For yet smaller volumes the bandwidth becomes
less well defined as the bands mix more strongly with the conduction
states and broaden.  The minority PDOS lies in the midst of Gd
$5d$ bands and is considerably broader down to $a/a_o$=0.70,
beyond which the difference becomes less noticeable.

\begin{figure}[tbp]
%\rotatebox{-90}
{\resizebox{7.7cm}{7.7cm}{\includegraphics{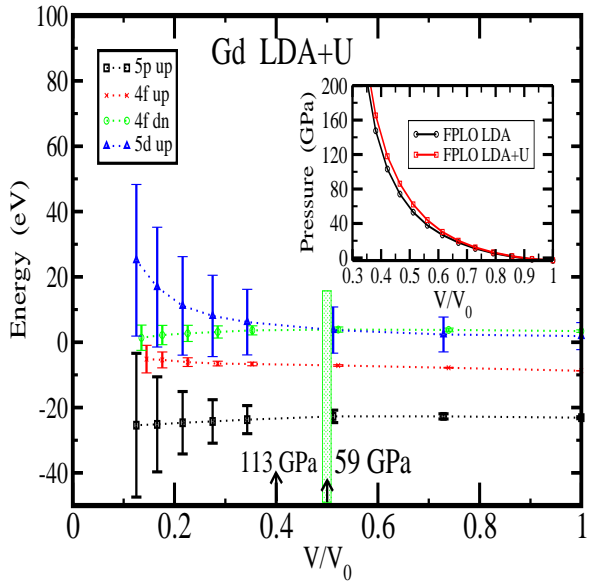}}}
\caption{(Color online)
Plot of band positions (lines) and widths (bars) of the
majority and minority $4f$ states, the semicore $5p$ bands,
and the valence $5d$ bands, for ferromagnetic Gd.  The bar at
$V/V_o$=0.5 ($\sim$59 GPa) marks the observed volume collapse 
transition, while the arrow at 113 GPa denotes the highest pressure
achieved so far in experiment.  These results were obtained from LDA+U
method, with $U$ varying with volume as given by
McMahan {\it et al}\cite{andy1}.
   }
\label{EandW}
\end{figure}

The position of the $4f$ states relative to the semicore $5p$,
and conduction $5d$ states, and their evolution with volume,
are pictured in Fig. \ref{EandW}.  The semicore $5p$ bands 
broaden to $\sim$10 eV by 200 GPa, but it requires supra-TPa
pressures to broaden them into the range of the majority $4f$
states.  The upturn in the log$P$ vs. $V$ curve in Fig. \ref{logPV}
in the vicinity of 100-200 GPa is probably due to $5p$ semicore overlap on
neighboring atoms (repulsion of closed shells as they come into
contact).  The $5d$ bands broaden in the standard way under
pressure, and begin to rise noticeably with respect to the $4f$
states beyond 60 GPa.

The minority $4f$ bands fall somewhat with respect to E$_F$ as
they broaden, both effects contributing to an increase in the minority
$4f$ occupation at the expense of $5d$ and $6sp$ character.
Since the majority $4f$ states remain full, the effect is that
the total $f$ count increases and the spin moment 
decreases (as discussed above).

The volume dependence of the $4f$ bandwidth in nonmagnetic Gd has been looked
at previously by McMahan {\it et al.}\cite{andy1}  They identified
the intrinsic width $W_{ff}$ from the bonding and antibonding 
values of the $4f$ logarithmic derivative;  $W_{ff}$ lies midway
(roughly halfway) between our majority and minority bandwidths,
see Fig. \ref{EandW}.  McMahan {\it et al.} also obtained a
hybridization contribution to the $4f$ width; both of these
would be included in our identified widths.  
Our widths, obtained for ferromagnetically ordered Gd, are difficult
to compare quantitatively with those of McMahan {\it et al.}, 
because the positions of our minority and majority states differ 
by 12 eV at P=0, decreasing under pressure.
Note that our minority and majority widths, 
obtained visually from
Fig. \ref{4fPDOS} differ by a factor of $\sim 6$ at $v$=1.0,
still by a factor of 2.5 at $v$=0.3, and only become equal in the
$v<0.2$ range.  

\subsection{Comments on Mott Transition}
In the simplest picture (single-band Hubbard model) the Mott
transition 
is controlled by the competition of
kinetic ($W$) and potential ($U$) energies, with the transition
occurring around $W \approx U$.  This transition is normally
pictured as a simultaneous insulator-to-metal, moment collapse, 
and presumably
also volume collapse transition.  In Gd, however, no change in moment is
observed\cite{maddox} across the volume collapse transition at 59 GPa.

\begin{figure}[tbp]
%\rotatebox{-90}
{\resizebox{7.7cm}{7.7cm}{\includegraphics{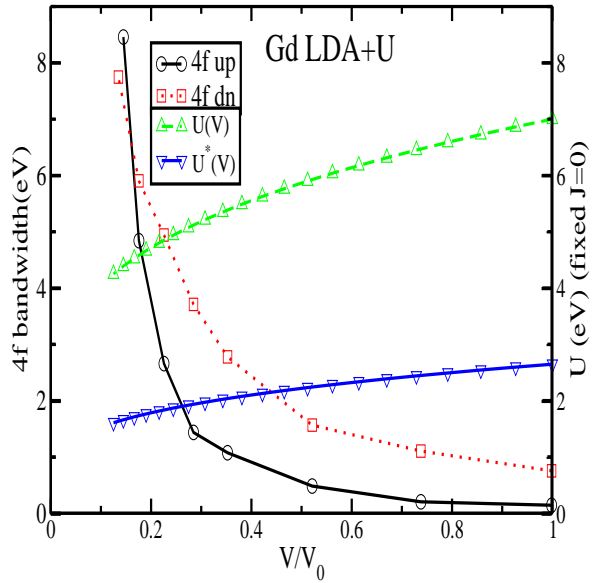}}}
\caption{(Color online)
Plot of the $4f$ bandwidths (both majority and minority),
together with the volume-dependent Coulomb repulsion U
from McMahan.\cite{andy1}  The simple crossover criterion
$W_f \approx U$ occurs around $V/V_o = 0.20-0.25$, corresponding
roughly to a pressure of 700-1000 GPa.  Also pictured is
$U^* \equiv U/\sqrt{7}$, see text for discussion.
}
\label{WvsU}
\end{figure}

In Fig. \ref{WvsU} the $4f$
bandwidths (majority and minority) and the Coulomb $U$
of McMahan {\it et al.}\cite{andy1} are plotted versus volume.  The
region $W \approx U$ occurs around $v\sim$0.20-0.25. This
volume corresponds to a calculated pressure in the
general neighborhood of 1 TPa, indicative of an extremely stable
moment well beyond present capabilities of static pressure cells.
This criterion however presumes a simple single band system,
which Gd is not.

Gunnarsson, Koch, and Martin have considered the Mott transition 
in the multiband Hubbard model,\cite{OG1,OG2,OG3}
and found that the additional channels for hopping
favored kinetic processes that reduced the effect of the 
Coulomb repulsion.  They argued that the criterion involved the
inverse square root of the degeneracy, which can be characterized
by an effective repulsion $U^* = U/\sqrt{7}$
(for $f$ states the degeneracy
is $2\ell$ +1 = 7).
The Mott transition could then be anticipated
in the range $W \approx U^*$, for which $U^*(V)$ has also been
included in Fig. \ref{WvsU}.  Taking W as the average of the majority
and minority widths gives the crossover around 
$v_c\sim$0.35 ($P_c\sim$ 200 GPa);
taking W more realistically as the majority bandwidth gives 
$v_c\sim$0.25 ($P_c\sim$ 750 GPa).

Another viewpoint on the ``Mott transition'' in the $4f$ system is
that it can be identified with the `metallization' of the
$4f$ bands, which might be expected to be where the occupied and 
unoccupied bands overlap.
These are respectively the majority and minority bands.  Significant
overlap occurs only above 2 TPa ($v <$ 0.20) in Fig. \ref{WvsU}.
The fact is that metallization (however defined for a $4f$ system
in the midst of uncorrelated itinerant conduction bands) and
moment collapse need not coincide, and the concept of Mott
transition may need to be generalized.

\section{Summary}
In this paper we have applied the correlated band theory LDA+U
method to probe the electronic and magnetic character of elemental
Gd under pressure.  The calculated moment decreases slowly 
down to $V/V_o$ = 0.20 ($P >$ 1 TPa), and only at smaller volumes
does the moment decrease more rapidly.  Still, no identifiable
moment collapse has been obtained.  Metallization, defined as
overlap of unoccupied with occupied bands, also does not occur 
until the same range of volume/pressure.  However, information from
studies of the multiband Hubbard model, and comparison of the
bandwidth to $U/\sqrt{N}$ ratio ($N$=7 is the $4f$ degeneracy)
suggests a ``Mott transition'' might be expected in the broad vicinity of
500 GPa.  

The same LDA+U method, and three different correlated band methods
have been applied to antiferromagnetic MnO.  The manganese configuration
is half-filled and fully polarized, as is Gd, with the difference
being that it is $3d$ and an antiferromagnetic insulator rather than 
$4f$ in a background of itinerant bands.
All methods obtained a volume
collapse from a high-spin to low-spin configuration.  Surprisingly,
the collapse was not to nonmagnetic but rather to a spin-half result.

The critical pressures for transitions suggested by the present
study (minimum of 200 GPa, more likely around 750 GPa) lie 
well above the volume collapse transition that
is observed at 59 GPa.  At this point in our understanding of the
$4f$ shell in Gd, there seems to be no viable alternative
to the suggestion by Maddox {\it et al.} that Gd provides an example
of the Kondo volume collapse mechanism.\cite{maddox}

\section{Acknowledgments}
This work has benefited greatly from a number of exchanges of
information and ideas with A. K. McMahan.
We have profited from many
discussions on Gd with C. S. Yoo, B. Maddox, R. T. Scalettar, and A.
Lazicki, and on the moment collapse question with M. D.
Johannes and J. Kune\v{s}.  We thank M. D. Johannes and R. T. Scalettar
for a careful
reading of the manuscript.
Support from the Alexander von Humboldt Foundation,
and the hospitality of IFW Dresden, during the 
preparation of this manuscript is 
gratefully acknowledged.
This work was supported by Department of Energy grant DE-FG03-01ER45876,
by Strategic Science Academic Alliance Program grant DE-FG03-03NA00071,
and by the DOE Computational Materials Science Network.


\begin{thebibliography}{10}

\bibitem{andy1}A. K. McMahan, C. Huscroft,
  R. T. Scalettar, and E. L. Pollock,
  J. Computer-Aided Materials Design {\bf 5}, 131 (1998);
  cond-mat/9805064.
\bibitem{err}D. Errandonea, R. Boehler, and M. Ross,
    Phys. Rev. Lett. {\bf 85}, 3444 (2000).
\bibitem{borje}B. Johansson, Philos. Mag. {\bf 30}, 30 (1974).
\bibitem{allen}J. W. Allen and R. M. Martin, Phys. Rev. Lett.
  {\bf 49}, 1106 (1982).
\bibitem{lavagna}M. Lavagna, C. Lacroix, and M. Cyrot, Phys.
  Lett. {\bf 90}A, 210 (1982); J. Phys. F {\bf 13}, 1008 (1983).

\bibitem{hua}H. Hua, Y. K. Vohra, J. Akella, S. T. Weir, R.
  Ahuja, and B. Johansson, Rev. High Pressure
  Sci. Technol. {\bf 7}, 233 (1988).
\bibitem{akella}J. Akella, G. S. Smith, and A. P. Jephcoat,
  J. Phys. Chem. Solids {\bf 49}, 573 (1988).
\bibitem{maddox} B. R. Maddox, A. Lazicki, C. S. Yoo, V. Iota,
  M. Chen, A. K. McMahan, M. Y. Hu, P. Chow, R. T. Scalettar,
  and W. E. Pickett, Phys. Rev. Lett. {\bf 96}, 256403 (2006).

\bibitem{kondo1}J. W. Allen and R. M. Martin, Phys. Rev. Lett.
  {\bf 49}, 1106 (1982).
\bibitem{kondo2}M. Lavanga, C. Lacroix, and M. Cyrot, Phys. Lett.
  {\bf 90A}, 210 (1982).

\bibitem{held}K. Held, A. K. McMahan, and R. T. Scalettar,
  Phys. Rev. Lett. {\bf 87}, 276404 (2001).
\bibitem{andy2}A. K. McMahan, Phys. Rev. B {\bf 72}, 115125 (2005).

\bibitem{SIC}A. Svane, W. M. Temmerman, Z. Szotek, J. Laegsgaard,
  and H. Winter, Intl. J. Quant. Chem. {\bf 77}, 799 (2000).
\bibitem{MnO}D. Kasinathan, J. Kunes, K. Koepernik, C. V. Diaconu, 
  R. L. Martin, I. Prodan, C. E. Scuseria, N. Spaldin, L. Petit, 
  T. C. Schulthess, and W. E. Pickett, cond-mat/0605430. 

\bibitem{johansson}B. Johansson and A. Rosengren, Phys. Rev. B
  {\bf 11}, 2836 (1975).

\bibitem{FPLO} K.Koepernik and H.Eschrig, Phys. Rev. B {\bf 59}, 1743 (1999).
\bibitem{PW92}J. P. Perdew and Y. Wang, Phys. Rev. B {\bf 45}, 13244 (1992).
\bibitem{WIEN2k}K. Schwarz, P. Blaha and G. K. H. Madsen, 
  Comput. Phys. Commun. {\bf 147}. 71 (2002).

\bibitem{robinson}L. B. Robinson, F. Milstein, and A. Jayaraman,
  Phys. Rev. {\bf 134}, A187 (1964).
\bibitem{jackson1}D. D. Jackson, C. Aracne-Ruddle, V. Malba, S. T. Weir,
  S. A. Catledge, and Y. K. Vohra, Rev. Sci. Instr. {\bf 74}, 2467 (2003).
\bibitem{jackson2}D. D. Jackson, V. Malba, S. T. Weir,
  P. A. Baker, and Y. K. Vohra, Phys. Rev. B {\bf 71}, 184416 (2005).
\bibitem{kurz}P. Kurz, B. Bihlmayer, and S. Bl\"ugel, J. Phys.:
  Condens. Matt. {\bf 14}, 6353 (2002).

\bibitem{sasha}A. B. Shick, A. I. Liechtenstein, and W. E. Pickett,
  Phys. Rev. B {\bf 60}, 10763 (1999). 
\bibitem{EuN}M. D. Johannes and W. E. Pickett, Phys. Rev. B
  {\bf 72}, 195116 (2005).
\bibitem{shick}A. B. Shick, W. E. Pickett, and C. S. Fadley,
  J. Appl. Phys. {\bf 87}, 5878 (2000).

\bibitem{OG1}O. Gunnarsson, E. Koch, and R. M. Martin, Phys. Rev. B {\bf 54},
  R11026 (1996).
\bibitem{OG2}O. Gunnarsson, E. Koch, and R. M. Martin, Phys. Rev. B {\bf 56},
  1146 (1997).
\bibitem{OG3}E. Koch, O. Gunnarsson, and R. M. Martin, Comp. Phys. Commun.
  {\bf 127}, 137 (2000).

\end{thebibliography}
\end{document}